\documentclass{emulateapj}
\usepackage{graphicx}
\begin{document}

\def\etal{et al.\ \rm}
\def\ba{\begin{eqnarray}}
\def\ea{\end{eqnarray}}
\def\etal{et al.\ \rm}

\title{Constraint on the giant planet production by core accretion.}

\author{Roman R. Rafikov\altaffilmark{1,2}}
\altaffiltext{1}{Department of Astrophysical Sciences, 
Princeton University, Ivy Lane, Princeton, NJ 08540; 
rrr@astro.princeton.edu}
\altaffiltext{2}{Sloan Fellow}


\begin{abstract}
The issue of giant planet formation by core instability (CI) far 
from the central star is rather controversial because the 
growth of massive solid core necessary for triggering the CI can 
take longer than the lifetime of the  
protoplanetary disk. In this work we assess the range of separations
at which the CI may operate by (1) allowing for arbitrary (physically 
meaningful) rate of planetesimal accretion by the core and (2) 
properly taking into account the dependence of the critical mass 
for the CI on the 
planetesimal accretion luminosity. This self-consistent approach 
distinguishes our work from similar studies in which only a 
specific planetesimal accretion regime was explored and/or the 
critical core mass was fixed at some arbitrary level. 
We demonstrate that the largest separation at which the CI can 
occur within $3$ Myr corresponds to the surface density of solids 
in the disk $\gtrsim 0.1$ g cm$^{-2}$ and is 
$40-50$ AU in the minimum mass Solar nebula. 
This limiting separation is achieved when 
the planetesimal accretion proceeds at the fastest possible rate, 
even though the high associated accretion luminosity increases
the critical core mass delaying the onset of the CI. 
Our constraints are independent of the mass of the central star 
and vary only weakly with the core density and its atmospheric 
opacity. We also discuss various factors which can strengthen or 
weaken our limits on the operation of the CI. 

\end{abstract}



\section{Introduction.}  
\label{sect:intro}

Recent discoveries of distant planetary companions (separations of
at least tens of AU) around several nearby stars by direct imaging 
(Marois \etal 2008; Thalmann \etal 2009) 
have stimulated investigation of their formation mechanisms.
Two contending theories of the planet formation --- the core instability 
(Perri \& Cameron 1974; Harris 1978; Mizuno 1980; hereafter CI) 
and the gravitational 
instability (Cameron 1978; Boss 1998) --- differ quite dramatically in
their ability to form planets at various distances from the star.
 
It is generally thought that CI should be capable of 
forming giant planets close to the star, at separations 
$\lesssim 10$ AU. Both theoretical modelling (Mizuno 1980; 
Stevenson 1982; Pollack \etal 1996) and the existence
of two gas giants at semi-major axes of $5.2$ AU and $9.5$ AU in 
our own Solar System (coupled with some knowledge about the 
evolution of the planetary orbital architecture in the early Solar 
System) attest to this statement. At the same time it seems 
unlikely that giant planets can be formed by the gravitational 
instability in the inner parts of protoplanetary disks because 
of the long local cooling time of gas (Matzner \& Levin 2005; 
Rafikov 2005, 2007). On the contrary, giant planet formation by 
the direct gravitational instability appears feasible at large 
distances from the star (Boley 2009; Rafikov 2009; Clarke 2009; 
but see Boss (2006) for an opposite view) where the disk cooling 
time is short enough to permit fragmentation of the gravitationally 
unstable disk (Gammie 2001). 
Whether the CI can operate beyond several tens of AU 
from the star is not so clear. 

The well known problem faced by the CI far from the star is that
the buildup of massive refractory core (necessary for triggering 
the vigorous gas accretion) is thought to take very long time,
longer than the $1-10$ Myr lifetime of the gaseous component of 
protoplanetary nebula. Studies illustrating this problem are 
usually based on two key assumptions (Dodson-Robinson \etal 2009): 
first, that the CI is triggered whenever a 
$M_{crit}\sim 10$ M$_\oplus$ core is built by planetesimal 
accretion, and, second, that the planetesimals accrete onto the growing
core at rather modest rates (Ida \& Lin 2004; 
we will explain later what do we mean
by that). When these conditions are imposed it is generally
found that the CI does not commence at $a\gtrsim 20-30$ AU prior 
to the nebular gas removal.

In this note we show that aforementioned assumptions are too restrictive and 
neither of them need to be adopted in determining the feasibility 
of the CI. First, planetesimal accretion by the growing core can 
(at least potentially) be much faster than what has been previously 
assumed. Second, the critical core mass itself strongly depends 
on the planetesimal accretion rate. Based on these observations we 
formulate in this work a novel constraint on the operation of the 
CI in protoplanetary disks.

\section{Planetesimal accretion.}
\label{sect:accr}

Onset of the CI is intimately related to the accretion of 
solid material by the protoplanetary core, as discussed in 
more detail in \S \ref{sect:atm}. For this reason we start 
by reviewing the process of planetesimal accretion by growing cores.

It is well known that the rate at which a core accretes 
planetesimals is a strong function of planetesimal random 
velocities (Dones \& Tremaine 1993; hereafter DT93). 
An important parameter for determining the dynamical state of 
planetesimal population is the ratio $p=R_c/R_H$ of the core 
radius $R_c$ to the Hill radius $R_H\equiv a(M_c/M_\star)^{1/3}$, 
where $a$ and $M_c$ are the semi-major axis and the mass of 
the core, and $M_\star$ is the mass of the star.
At large separations from the star
\ba
p=\left(\frac{3M_\star}{4\pi \rho_c a^3}\right)^{1/3}
\approx 5\times 10^{-4}a_{10}^{-1}
M_{\star,1}^{1/3}\rho_{c,1}^{-1/3}
\label{eq:p}
\ea
is much less than unity, where $\rho_c$ is the core density 
and we define $a_{10}\equiv a/(10~\mbox{AU})$, 
$M_{\star,1}\equiv M_\star/M_\odot$, $\rho_{c,1}=\rho/1$ g cm$^{-3}$. 

DT93 have demonstrated that whenever $p\ll 1$  
there are four possible regimes of planetesimal accretion. 
\begin{itemize}
\item When the planetesimal random velocity dispersion\footnote{For 
simplicity here we do not differentiate between the horizontal
and vertical velocity dispersion.} 
$\sigma\gtrsim \Omega R_H p^{-1/2}$ ($\Omega=\sqrt{GM_\star/a^3}$ 
is the angular frequency at the core's location) -- a regime
that we call {\it very high dispersion} -- accretion 
is very slow because gravitational focussing is inefficient and 
the collision cross-section is given by the geometric 
cross-section of the core. 
\item When 
$\Omega R_H\lesssim \sigma\lesssim \Omega R_H p^{-1/2}$ 
({\it high dispersion} in the notation of DT93) the 
relative planetesimals velocity with respect to the core 
$v_{rel}\sim\sigma$. Gravitational focussing is  
important and planetesimal accretion rate $\dot M$ grows as
$\sigma$ goes down. 
\item When 
$\Omega R_H p^{1/2}\lesssim \sigma\lesssim \Omega R_H$ 
({\it intermediate dispersion})
one finds that $v_{rel}$ is no longer determined by the random 
motions of planetesimals but is rather set by the shear in
the differentially rotating disk, so that $v_{rel}\sim\Omega R_H$.
Gravitational focussing saturates at a constant value whenever
$\sigma\lesssim \Omega R_H$ and $\dot M$ keeps increasing as $\sigma$
decreases simply because the thickness of the disk goes down increasing
local volume density of accreting bodies. 
\item Finally, when 
$\sigma\lesssim\Omega R_H p^{1/2}$ ({\it very low dispersion})
disk becomes so thin that a core can accrete the whole vertical 
column of material doomed for collision with it. Given that 
focussing is at its maximum this situation corresponds to the 
most efficient accretion regime. 
\end{itemize}

The rate of accretion in
the very low dispersion case is (DT93)
\ba
&& \dot M=\dot M_{max}\approx (6.47\pm 0.02)\Omega p^{1/2}
\Sigma_s R_H^2,
\label{eq:max_rate}\\
&& \approx 5.1~G^{1/2}M_c^{2/3}\Sigma_s\rho_c^{-1/6},
\ea
where $\Sigma_s$ is the surface density of solids. In this study we 
parametrize $\Sigma_s$ as 
\ba
\Sigma_s=1\Sigma_1 a_{10}^{-\alpha}\,\mbox{g cm}^{-2},
\label{eq:Sig_s}
\ea
where $\Sigma_1=\Sigma_s/(1$ g cm$^{-2})$ and $\alpha$ is a constant 
(minimum mass Solar nebula (MMSN)
corresponds to $\alpha\approx 3/2$ and $\Sigma_1\approx 1$).
Note that $\dot M_{max}$ is independent of $M_\star$.

For comparison, it is often assumed that protoplanetary cores
predominantly accrete planetesimals with the dispersion of 
random velocities $\sigma=\Omega R_H$, which corresponds to the 
{\it transition} between the so called shear- and dispersion-dominated 
dynamical regimes (e.g. Dodson-Robinson 2009). This is what we 
called a ``modest'' accretion rate before. At this transition 
DT93 find 
\ba
\dot M_{tr}\approx (6-10)p\Sigma_s\Omega R_H^2, 
\label{eq:M_tr}
\ea
and one immediately sees that $\dot M_{tr}\ll\dot M_{max}$ since
$p\ll 1$.

Previous studies of the CI have been limited to considering 
planetesimal accretion only in the high-velocity regime 
($\sigma\gtrsim \Omega R_H$) or at the boundary with the 
intermediate velocity regime (i.e. for $\sigma\approx\Omega R_H$). 
While large (tens to hundreds km in 
size) planetesimals indeed likely accrete in these dynamical 
regimes at rather slow rates it is quite possible that most 
of the mass growth of the core is not due to accretion of these 
large bodies. According to Rafikov (2004a) and Goldreich \etal 
(2004), as the core becomes massive enough it dynamically 
excites large planetesimals in its vicinity since damping agents
such as gas drag are not effective for massive bodies.
This leads to fragmentation (rather than growth) of planetesimals 
when they collide with each other. Resulting 
fragmentation cascade converts significant fraction of the 
solid mass into small mass debris. These fragments finally reach 
small enough sizes that they are dynamically ``cooled'' by gas
drag to low velocities, which makes it possible for them to be accreted
at very high rate compatible with
$\dot M_{max}$ (Rafikov 2004a,b) in the very low dispersion regime. 
Recent coagulation calculations 
done by Kenyon \& Bromley (2009) confirm this general picture,
lending support to the possibility of maximally efficient accretion 
in the very low dispersion regime. 

In this work we are interested in obtaining a robust limit on 
the operation of core instability, which means that we need
to consider all possible modes of planetesimal accretion and 
actually {\it determine} the accretion regime facilitating the 
onset of the CI the most (instead of selecting it ad hoc). It 
is not obvious a priori which accretion regime is best for 
achieving the CI within the limited amount of time: on one
hand high $\dot M$ allows to build the core of a given mass 
faster, but on the other hand high $\dot M$ also implies 
larger critical mass (see \S \ref{sect:mcrit}), which takes 
longer to build.

We will discover in \S \ref{sect:mcrit} that fastest route 
to the CI lies through the most efficient accretion at the rate 
$\sim \dot M_{max}$ and for this reason we parametrize 
planetesimal accretion rate as 
\ba
\dot M &=&\chi\dot M_{max}\approx 6.47\chi p^{1/2}
\Sigma_s\Omega a^2\left(\frac{M_c}{M_\star}\right)^{2/3},
\label{eq:mdot}\\
&\approx & 2\times 10^{-5}\mbox{M}_\oplus~\mbox{yr}^{-1}\chi
a_{10}^{-\alpha}\Sigma_1\rho_{c,1}^{-1/6}M_{c,10}^{2/3},\nonumber
\ea
($M_{c,10}\equiv M_c/10$ M$_\oplus$) where the parameter 
$\chi$ accounts for the deviation of $\dot M$ from 
$\dot M_{max}$. In general $\chi$ may be a function of $M_c$
and $a$, and it is expected that $\chi\lesssim 1$
(although in some situations $\chi\gtrsim 1$ may possible, 
see \S \ref{sect:disc}). But we will show in \S \ref{sect:mcrit}
that slow accretion (i.e. $\chi\ll 1$) results in 
smaller distance from the star at which planets could form
by the CI than in the case $\chi\sim 1$. It should also be
emphasized that our concentration on the highest $\dot M$ 
accretion regime is not only for the sake of the argument -- 
as mentioned before this regime may actually occur quite 
naturally in the course of the core buildup.

Accretion luminosity corresponding to the $\dot M$ given 
by equation (\ref{eq:mdot}) is 
\ba
L &=& \frac{GM_c\dot M}{R_c}\approx 8.2\chi\Sigma_s G^{3/2} M_c^{4/3}
\rho_c^{1/6}
\label{eq:lumin}\\
& \approx & 8\times 10^{-6}\mbox{L}_\odot ~\chi
a_{10}^{-\alpha}\Sigma_1\rho_{c,1}^{1/6}M_{c,10}^{4/3}.\nonumber
\ea

According to equation (\ref{eq:mdot}) accretion at the 
rate $\dot M_{max}$ results in the core radius growing linearly 
with time\footnote{One can easily see that accretion at the rate 
$\dot M_{tr}$ (see Eq. [\ref{eq:M_tr}]) also leads to 
$M_c^{1/3}\propto \tau$, although the
coefficient of this relation is considerably smaller than in
the case of $\dot M=\dot M_{max}$ and has different scaling with 
$a$.}. Time needed for the core mass to reach a predetermined 
value $M_c$ by accretion at the rate $\chi\dot M_{max}$ is 
[see equation (\ref{eq:max_rate})] 
\ba
\tau(M_c)\approx 0.6\frac{\rho_c^{1/6}M_c^{1/3}}{\chi\Sigma_s G^{1/2}}
\approx 0.3~\mbox{Myr}~a_{10}^\alpha 
\frac{\rho_{c,1}^{1/6}M_{c,10}^{1/3}}{\chi\Sigma_1}.
\label{eq:timescale}
\ea
It follows from this formula that a core can grow to 10 M$_\oplus$ 
in 3 Myr by accretion at the rate $\dot M_{max}$ even at $\sim 50$ 
AU in the MMSN. Previous results which did not account for the 
factors leading to the fast accretion (planetesimal fragmentation 
and gas drag on debris) predict much smaller distance 
($\lesssim 20$ AU) at which such core would be able to form on
that time scale. 

Having said all that, in the next section we show that the 
formation of the 10 M$_\oplus$ core is not a prerequisite 
for a CI and that the $M_{crit}$ is a function of accretion 
rate.

\section{Protoplanetary atmosphere.}
\label{sect:atm}

Numerical calculations (Mizuno 1980; Pollack \etal 1996; 
Ikoma \etal 2000) and analytical theory (Stevenson 1982) show 
that CI commences when the core mass $M_{crit}$ is 
so large that the mass of its atmosphere is comparable 
to the core mass itself, 
\ba
M_{atm}(M_{crit})=\eta M_{crit},
\label{eq:instab_cond}
\ea 
i.e. when the gaseous component of the protoplanet becomes 
self-gravitating. Here $\eta$ is a parameter of order unity; 
its exact value may depend on the accretion history of the core
(Ikoma \etal 2000). Condition (\ref{eq:instab_cond}) should be 
regarded as an equation for $M_{crit}$ which can be easily solved 
as long as the dependence of $M_{atm}$ on $M_c$ and other 
parameters of the problem is specified. Thus, calculation of 
$M_{crit}$ involves understanding properties of planetary 
atmospheres and calculation of $M_{atm}$ in particular.

It has been known since the works of Mizuno (1980) and 
Stevenson (1982) that the critical core mass is a rather 
weak function of the density $\rho_0$ and temperature $T_0$ 
of the surrounding nebula (see Rafikov [2006] and \S 
\ref{subsect:opacity} for the discussion of the conditions 
under which this behavior is expected). This has led to the 
wide acceptance 
of the CI idea since the interiors of the Solar System giant 
planets are believed to harbor $\sim 10$ M$_\oplus$ cores 
despite their different separations from the Sun. 
Unfortunately, this observation 
of the $M_{crit}$ invariance has shadowed the fact first 
noticed by Stevenson (1982) that $M_{crit}$ is a strong function 
of the planetesimal accretion luminosity $L$, or, equivalently, 
planetesimal accretion rate $\dot M$. This fact has been 
subsequently confirmed both numerically (Ikoma \etal 2000; 
Hori \& Ikoma 2010) and analytically (Rafikov 2006).

Stevenson's analytical (1982) results strictly apply only to cores 
possessing radiative atmospheres with constant opacity. 
Rafikov (2006; hereafter R06) has studied more general types of 
atmospheres and has shown that they segregate into two
classes depending on whether $L$ is higher or lower than some 
critical luminosity $L_{cr}$.
Whenever $L\gtrsim L_{cr}$ the intense energy release near the
core surface makes protoplanetary atmosphere and the
nebular gas in the Hill sphere of the core convectively 
unstable. Such atmosphere have high entropy and rather 
low mass relative to the mass of the core. In the opposite
case, when $L\lesssim L_{cr}$, protoplanetary atmosphere is
separated from the nebular gas by a roughly isothermal shell 
of gas in which energy is transported radiatively and gas 
entropy decreases from the nebular value to a much smaller 
value characteristic for the atmosphere. Density in this 
shell increases roughly exponentially towards the planet 
which makes $M_{atm}$ much higher (for a given core 
mass) than in the high-luminosity case. 

It was demonstrated in R06 that far from the star 
protoplanetary atmospheres are virtually always characterized 
by $L\ll L_{cr}$, even if planetesimal accretion proceeds 
at the maximally efficient rate $\dot M_{max}$. For that 
reason we will consider only the low luminosity atmospheres in
this work.
The total atmospheric mass $M_{atm}$ in the low luminosity case 
was computed in R06 by making simplifying assumptions about 
the conditions in the deep layers of the planetary atmosphere:
either a polytropic model with constant polytropic index 
(mimicking the fully convective interior) or a fully radiative
atmosphere with a simple power-law parametrization of the 
opacity dependence on gas pressure and temperature.
Under these assumptions a significant fraction of the atmospheric 
mass resides in the outer\footnote{This is different from the case 
of an atmosphere with constant opacity considered by Stevenson 
(1982).}
layer, near the inner boundary of the roughly isothermal external 
radiative zone, in agreement with calculations by Mizuno (1980). 
It can then be shown that (R06)
\ba
M_{atm}\approx \zeta\left(\frac{G M_c\mu}{k}\right)^4
\frac{\sigma}{\kappa_0 L},
\label{eq:Matm}
\ea
where $k_B$ is the Boltzmann constant and $\zeta$ is a weakly 
varying factor, which can be computed
exactly for a given density distribution inside the atmosphere. 
This formula shows how $M_{atm}$ depends on important parameters 
like $M_c$ and $\kappa_0$ and on accretion history, to which 
$L$ is sensitive. 

However it is not obvious that the simplifying assumptions 
about the atmospheric properties 
used in deriving formula (\ref{eq:Matm}) are justified given
the complexity of the physical effects encountered deep in 
the atmosphere: self-gravity gradually becomes
important as $M_c$ approaches $M_{crit}$, grain sublimation changes
opacity in a non-trivial fashion, the equation of state 
may be varying with depth because of molecular dissociation,
an atmosphere may have both radiative and convective regions at
the same time. These complications likely do not affect the 
qualitative conclusions reached in R06 but should be important 
for quantitative comparisons. 

On the other hand, numerical calculations of the CI 
which include the aforementioned physical effects (Ikoma \etal 2000) 
are typically 
limited to exploring the dependence of $M_{crit}$ on only a 
limited set of input parameters such as $\kappa_0$ and $\dot M$.
The latter is usually taken to be constant through the 
calculation and this assumption significantly limits the direct 
applicability of these numerical results to realistic situations 
(since typically $\dot M$ increases as $M_c$ grows), including 
our present study. 

To be able to apply the existing numerical results for the 
cases when the core accretion history is non-trivial 
($\dot M\neq$ const) while at
the same time keeping the flexibility of the 
theory outlined in R06 we have resorted to the following approach.
First, we still use formula (\ref{eq:Matm}) to calculate $M_{atm}$
but now we do not assume coefficient $\zeta$ to be constant as
various physical effects in the deep atmosphere introduce 
additional variations of the density profile not captured by
the analytical theory of R06. Second, using equation 
(\ref{eq:instab_cond}) we calculate $M_{crit}$ under the 
assumption $\dot M=const$ used in Ikoma \etal (2000). With this 
information in hand we can calibrate the dependence of $\zeta$ 
on various physical parameters of the problem by comparing our
results with those of Ikoma \etal (2000). Finally, after $\zeta$
has been calibrated for a particular accretion history 
$\dot M=$const, equation (\ref{eq:Matm}) can be applied for 
a more general situation, e.g. for the $\dot M(M_c)$ dependence
used in this work.

This approach should work well as long as the state of the
atmosphere is fully determined by the current value of $\dot M$, 
and is independent of the full accretion 
history. For this to be the case the thermal timescale of the
atmosphere must be short compared to the planetesimal accretion 
timescale, and this has been verified in R06.  
This gives us confidence that the outlined method 
should be robust and justifies its application to the problem at hand. 

Since $\zeta$ is  
predominantly affected by the changes of the thermodynamical 
state of material in the atmosphere (grain sublimation, 
molecular dissociation, variations of 
the equation of state depend on gas temperature and density) one 
should expect $\zeta$ to depend most strongly on $M_c$ and $L$ 
--- variables that determine $T$ and $\rho$ deep in the envelope. 
Ikoma \etal (2000) have shown that in their case $M_{crit}$
scales with $\dot M$ and $\kappa_0$ roughly as power laws. 
This motivates us to look for the dependence of $\zeta$ on $M_c$
and $L$ also in the power law form. The details of the calibration 
procedure are presented in Appendix \ref{app:zeta} where we
show that 
\ba
&& \zeta= \zeta_0 \eta\, M_c^\delta,~~~\delta\approx 1.2,
\label{eq:zeta}\\
&& \zeta_0\approx  
4\times 10^{-32}\,\mbox{g}^{-\delta}\approx 82\,M_\oplus^{-\delta}\nonumber
\ea
provides a reasonably good fit to the numerical results of Ikoma 
\etal (2000). Note that $\zeta$ has been found to depend on $L$ 
only weakly, and can be considered a function of $M_c$ only.
In deriving this result for $\zeta$ we have assumed that
the coefficient $\eta$ appearing in the relation 
(\ref{eq:instab_cond}) is independent of the planetesimal accretion 
history of the core and is thus the same in Ikoma \etal work as well
as in our case.

\section{Critical core mass.}
\label{sect:mcrit}

Results obtained in the previous section allow us to compute 
$M_{crit}$ for $\dot M$ given by equation (\ref{eq:mdot}): 
substituting $M_{atm}$ in the form (\ref{eq:Matm}) 
with $\zeta$ given by equation (\ref{eq:zeta}) into the 
instability condition (\ref{eq:instab_cond}) and using equation 
(\ref{eq:lumin}) for core's accretion luminosity we find 
\ba
\label{eq:Mcrit}
M_{crit} & \approx & \left[\frac{8.2}{\zeta_0}
\left(\frac{k_B}{\mu}\right)^4\frac{\chi\Sigma_s\rho_c^{1/6}\kappa_0}
{\sigma G^{5/2}}\right]^{3/(5+3\delta)}\\
& \approx & 40~\mbox{M}_\oplus\left(
\frac{\chi\Sigma_1\kappa_{0.1}\rho_{c,1}^{1/6}}{\mu_2^4}\right)^{0.35}
a_{10}^{-0.35\alpha},\nonumber
\ea
where $\kappa_{0.1}\equiv \kappa_0/(0.1$ cm$^2$ g$^{-1})$ and 
$\mu_2\equiv \mu/(2m_H)$ ($m_H$ is the atomic hydrogen 
mass). 

Equation (\ref{eq:Mcrit}) shows among other things that $M_{crit}$ 
is indeed a function of the planetesimal accretion regime: 
according to equation (\ref{eq:mdot}) we may think of the 
free parameter $\chi$ as a direct measure of planetesimal 
$\dot M$, and $M_{crit}$ scales roughly as $\chi^{0.35}$. Just 
for illustration, if (as has been done in 
Dodson-Robinson \etal 2009) planetesimal random 
velocities were kept at the level $\sigma\sim \Omega R_H$ 
(rather than being essentially zero as assumed in deriving equation 
(\ref{eq:mdot})) the equation (\ref{eq:mdot}) could still be used 
but with $\chi\approx p^{1/2}\ll 1$ (DT93; R06). At $a=40$ AU where
$p\approx 10^{-4}$ one would then find $M_{crit}$ smaller by a 
factor of $5$ compared to what equation (\ref{eq:Mcrit}) predicts 
for $\chi=1$ at the same location. 

With equation (\ref{eq:Mcrit}) we also show for our chosen $\dot M$ 
behavior that $M_{crit}$ varies with the distance from the central 
object: for $\chi$=const the critical mass scales 
as $\approx a^{-0.53}$ in the MMSN ($\alpha=3/2$). 
This makes $M_{crit}$ at 10 AU 
twice as large as $M_{crit}$ at 40 AU, everything else being equal. 
Thus, taking $M_{crit}$ to be 
independent of the regime of planetesimal accretion and distance
from the central object is generally not justified.

\section{Limit on core instability.}
\label{sect:constr}

For the CI to happen before the nebular gas dispersal 
the growth time of the core with mass $M_{crit}$ must be less
than the lifetime of the protoplanetary nebula $\tau_{neb}$, i.e. 
$\tau(M_{crit})< \tau_{neb}$. Plugging in our result (\ref{eq:Mcrit}) 
into equation (\ref{eq:timescale}) this constraint can be rephrased in
terms of the {\it lower limit} on the surface density of solids: 
\ba
\chi\Sigma_s &>&\Sigma_{lim},\label{eq:sig_crit}\\
\Sigma_{lim}&=&0.74\left[
\frac{\rho_c^{(2+\delta)/2}\left(k_B/\mu\right)^4}
{\zeta_0\tau_{neb}^{5+3\delta}
G^{(10+3\delta)/2}}
\frac{\kappa_0}{\sigma}\right]^{1/(4+3\delta)}
\label{eq:Sig_lim}\\
& \approx & 0.1~\mbox{g sm}^{-2}\left(\frac{\tau_{neb}}{3\,\mbox{Myr}}
\right)^{-1.13}\frac{\rho_{c,1}^{0.21}\kappa_{0.1}^{0.13}}{
\mu_2^{0.53}}.\nonumber
\ea
This inequality represents the main result of this work --- a robust
lower limit on the {\it planetesimal} surface density at which
CI is capable of producing giant planets within a 
protoplanetary nebula lifetime. Note that $\Sigma_{lim}$ is a 
sensitive function of the nebula lifetime $\tau_{neb}$, while it 
depends rather weakly on the bulk density of the core $\rho_c$,
atmospheric dust opacity $\kappa_0$, and the mean molecular
weight of the atmospheric gas $\mu$. 

Now we can also determine the limiting core mass $M_{lim}$ defined
as the critical core mass for $\chi\Sigma_s=\Sigma_{lim}$, i.e. 
at the very extreme of the region where CI can still occur
within $\tau_{neb}$. It is found by substituting (\ref{eq:Sig_lim}) 
into (\ref{eq:Mcrit}):
\ba
M_{lim}&=&\left[\frac{4.9}{\zeta_0}
\left(\frac{k_B}{\mu}\right)^4\frac{\rho_c^{1/3}\kappa_0}
{\sigma G^{3}\tau_{neb}}\right]^{3/(4+3\delta)}
\label{eq:Mlim}\\
& \approx & 15~\mbox{M}_\oplus\,
\left(\frac{\tau_{neb}}{3~\mbox{Myr}}\right)^{-0.4}
\frac{\rho_{c,1}^{0.13}\kappa_{0.1}^{0.4}}{\mu_2^{1.58}}.\nonumber
\ea
By construction $M_{lim}$ is independent of the surface density
profile in the nebula. 
This mass is to be compared with the isolation mass $M_{iso}$ 
(the core mass
at which it has accreted all solid mass within its feeding zone) --- 
an annulus centered on core's orbit and having a full width equal 
to $\xi R_H$: 
\ba
M_{iso}&=&\frac{\left(2\pi\xi\Sigma_s a^2\right)^{3/2}}{M_\star^{1/2}},
\label{eq:M_iso}\\
& \approx &2\,\left(\xi/5\right)^{3/2}\Sigma_1^{3/2}a_{10}^{3/4}
\,M_\oplus\nonumber
\ea
(an MMSN-like density profile was used in making numerical estimate).
At $44$ AU one finds $M_{iso}\approx 6$ M$_\oplus$, which is smaller
than $M_{lim}$. However, a modest radial displacement of the core due 
to some type of migration can easily expose it to additional fresh 
material allowing $M_c$ to reach $M_{lim}$ (Alibert \etal 
2005). Alternatively, increasing $\Sigma_s$ (boosting up $\Sigma_1$) 
by a factor of 2 makes $M_{iso}\approx M_{lim}$ at 44 AU.

Quite interestingly, the value of $M_{lim}$
is not too far from $10$ M$_\oplus$ commonly accepted as the core
mass throughout the whole protoplanetary disk. This coincidence is 
accidental since our estimate of $M_{lim}$ was derived 
self-consistently rather than postulated in an ad hoc fashion.

\subsection{Limiting distance for core instability.}
\label{subsect:dist}

Given a constraint (\ref{eq:Sig_lim}) and having a particular 
model of the radial distribution of $\Sigma_s$ one can determine 
the maximal radial extent $a_{lim}$ of the region in protoplanetary disk 
in which the CI can produce giant planets within the nebula life 
time $\tau_{neb}$. Since $\Sigma_s$ is expected to be a decreasing 
function of $a$ this would only be possible for $a<a_{lim}$. 
Using our power-law parametrization (\ref{eq:Sig_s}) of $\Sigma_s$ 
we find that in a MMSN-like disk with $\alpha=3/2$
\ba
a_{lim}^{MMSN}\approx 44~\mbox{AU}~\frac{\left(\chi\Sigma_1
\right)^{2/3}\mu_2^{0.35}}{\rho_{c,1}^{0.14}\kappa_{0.1}^{0.09}}
\left(\frac{\tau_{neb}}{3~\mbox{Myr}}\right)^{0.75}.
\label{eq:a_lim}
\ea
Accretion at $\dot M=\dot M_{max}$ corresponds to $\chi=1$
in this formula. The limiting distance found in equation 
(\ref{eq:a_lim}) is similar to previous estimates 
(e.g. Dodson-Robinson \etal 2009) obtained for less vigorous 
planetesimal accretion (i.e. for smaller $\dot M$) and fixed 
$\dot M_{crit}=10$ M$_\oplus$ but this is just a coincidence. 

Equation (\ref{eq:a_lim}) shows that as $\Sigma_s$ (or $\Sigma_1$) 
increases, the extent of CI-capable part of the protoplanetary 
disk also grows. But $\Sigma_s$ cannot be increased without limit ---
at some point the gaseous component of the protoplanetary nebula
 would become self-gravitating,
and, depending on its cooling time (Gammie 2000), would either 
fragment or evolve quasi-viscously while maintaining the marginally 
gravitationally unstable state (Rafikov 2009; Clarke 2009). 
Thus the most extreme value of $a_{lim}$ can be obtained by taking
$\Sigma_s(a)=f_{dg}\Sigma_{Q=1}(a)$ in equation (\ref{eq:Sig_lim}), 
where $f_{dg}$ is the dust to gas ratio, which we take to
be $10^{-2}$, and $\Sigma_{Q=1}(a)$ is the gas surface density 
at which the disk is marginally gravitationally unstable, which 
happens when the Toomre $Q\equiv\Omega c_{s,0}/(\pi G\Sigma_{Q=1})$
is of order unity (Safronov 1960; Toomre 1964). Assuming that the 
disk is heated by a central star with luminosity 1 
$L_\odot$ at normal incidence (i.e. neglecting complications related 
to flaring geometry, Chiang \& Goldreich [1997]) we find that
\ba
f_{dg}\Sigma_{Q=1}(a)\approx 20\,a_{10}^{-7/4}\,\mbox{g cm}^{-2}.
\label{eq:Sigma_sg}
\ea
Using this density profile in equation (\ref{eq:Sig_lim}) we find
\ba
a_{lim}^{Q=1}\approx 200~\mbox{AU}~\frac{\chi^{0.57}
\mu_2^{0.3}}{\rho_{c,1}^{0.12}\kappa_{0.1}^{0.07}}
\left(\frac{\tau_{neb}}{3~\mbox{Myr}}\right)^{0.64}.
\label{eq:a_lim_sg}
\ea
In practice this limit will hardly apply to real systems 
because self-gravitating disks with $Q\sim 1$ would 
not persist for the lifetime of the nebula, as we already
mentioned.

\subsection{Sensitivity to planetesimal accretion efficiency.}
\label{subsect:acc_efficiency}

Through our calculations we have retained in all formulae the 
parameter $\chi$ defined in equation (\ref{eq:mdot}), which 
characterizes the efficiency of planetesimal accretion. This
allows us to see how the variation of planetesimal $\dot M$
with respect to $\dot M_{max}$ affects the
planet formation by the CI. From equation (\ref{eq:Sig_lim})
we see that smaller $\chi$ (corresponding to less efficient
accretion) results in {\it higher} $\Sigma_s$ and {\it smaller} 
$a_{lim}$, see equations (\ref{eq:a_lim}), (\ref{eq:a_lim_sg}).

\begin{figure}
\plotone{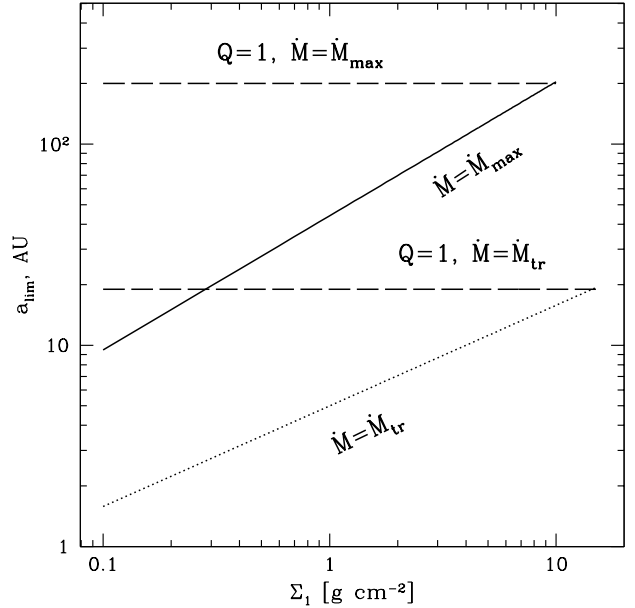}
\caption{
Plot of the limiting semi-major axis $a_{lim}$ beyond which core 
instability cannot produce planets within 3 Myr as a function of 
$\Sigma_1$ --- surface density of solids at 10 AU in a MMSN-like 
protoplanetary disk with $\Sigma_s\propto a^{-3/2}$ 
(see Eq. [\ref{eq:Sig_s}]). Solid curve is for planetesimal accretion 
proceeding at maximum possible rate $\dot M_{max}$, dotted curve
is for $\dot M=\dot M_{tr}$ (accretion of planetesimals with 
$\sigma\sim \Omega R_H$, see Eq. [\ref{eq:M_tr}]). Dashed lines
correspond to $a_{lim}$ in a 
marginally gravitationally unstable disk, both for 
$\dot M=\dot M_{max}$ (upper line) and 
$\dot M=\dot M_{tr}$ (lower line). See text for details.
\label{fig:f1}}
\end{figure}

As we emphasized in \S \ref{sect:atm}, this result is not trivial  
since less efficient accretion implies lower planetesimal 
luminosity, bigger $M_{atm}$ for the same $M_c$, and lower 
$M_{crit}$, making CI easier to get going. However, it turns 
out that the competing effect 
of being able to grow the massive core faster at higher $\dot M$
is more important for the CI to commence {\it within the 
limited amount of time}. For this reason the largest extent of
the region in which the CI can happen in time $\tau_{neb}$ is 
reached when the core is able to accrete at the highest possible 
rate, namely at $\dot M_{max}$ (we discuss whether it is potentially 
possible to {\it exceed} $\dot M_{max}$ in 
\S\ref{subsect:strengthen}). This makes our numerical estimates 
in equations (\ref{eq:a_lim}), (\ref{eq:a_lim_sg}) very robust.

Just for illustration let us also consider a situation in which 
the core accretes planetesimals at the rate $\dot M_{tr}$ given
by equation (\ref{eq:M_tr}), and determine $a_{lim}$ in this case. 
Comparing expressions (\ref{eq:max_rate}) and (\ref{eq:M_tr}) one 
can easily see that we can do this by simply setting 
$\chi=p^{1/2}$ ($p(a)$ is defined by equation [\ref{eq:p}]) either 
in equation  (\ref{eq:Sig_lim}) or in equation (\ref{eq:a_lim})
and determining  $a_{lim}$ from the resulting expression:
\ba
a_{lim}^{tr}\approx 5~\mbox{AU}~\frac{\Sigma_1^{0.5}
\mu_2^{0.27}M_{\star,1}^{0.08}}{\rho_{c,1}^{0.19}\kappa_{0.1}^{0.07}}
\left(\frac{\tau_{neb}}{3~\mbox{Myr}}\right)^{0.57},
\label{eq:a_lim_tr}
\ea
where we assumed MMSN-like disk properties to provide direct 
comparison with $a_{lim}^{MMSN}$ given by equation (\ref{eq:a_lim}).
If, as we did in deriving equation (\ref{eq:a_lim_sg}), we assume 
that the disk is marginally gravitationally unstable we would 
find $a_{lim}^{Q=1,tr}\approx 20$ AU for our standard choice of 
parameters. In Figure \ref{fig:f1}  
we display different expressions for $a_{lim}$ 
as functions of $\Sigma_1$ --- value of $\Sigma_s$ at 10 AU --- 
obtained for $\kappa_{0.1}=1$, $\rho_{c,1}=1$, $\mu_{2}=1$, 
$\tau_{neb}=3$ Myr under a variety of assumptions regarding $\Sigma_s$
and $\chi$.

Clearly, less efficient accretion results in significantly
more compact region of the disk where the CI can occur. In fact,
equation (\ref{eq:a_lim_tr}) implies that the formation of Saturn 
by the CI at $10$ AU in our Solar System would only be possible 
if the surface density in the proto-Solar nebula was at least 4 
times higher than in the MMSN or if the nebula dissipation 
timescale was at least 10 Myr, in agreement with existing
studies which assume planetesimal accretion at 
$\dot M\sim \dot M_{tr}$ (Ida \& Lin 2004).

\subsection{Sensitivity to opacity variations.}
\label{subsect:opacity}

Rather interestingly, our calculations find very weak dependence of 
both $\Sigma_{lim}$ and $a_{lim}$ on $\kappa_0$. Although the 
reduction of $\kappa_0$ does help to reduce $M_{crit}$ and 
$M_{lim}$, the values of $\Sigma_{lim}$ and $a_{lim}$ remain 
virtually unaffected. Given the large uncertainly in the
value of $\kappa_0$ this property further strengthens our 
estimates of $\Sigma_{lim}$ and $a_{lim}$.

Recently Hori \& Ikoma (2010) calculated
$M_{crit}$ as a function of $\dot M$ for protoplanets with 
dust-free (possible if dust grains sediment from the outer 
layers of the protoplanetary atmosphere) and metal-free 
(i.e. containing only H and He) atmospheres and found 
values of $M_{crit}$ lower by up to an order of magnitude 
(as low as $2$ M$_\odot$ in the metal-free case for 
$\dot M=10^{-6}$ M$_\odot$ yr$^{-1}$) than in the dusty 
case. It would certainly be interesting to 
repeat calculations done in \S\S\ref{sect:mcrit}, 
\ref{sect:constr} and Appendix \ref{app:zeta} for the case of 
dust-free atmosphere to see how this extreme reduction of opacity 
would extend the radial range available for CI.

In practice, we cannot do such calculation at present  
since it is not possible to calibrate $M_{atm}$ against the results 
of Hori \& Ikoma (2010) as we did in equations 
(\ref{eq:Matm})-(\ref{eq:zeta}) and Appendix \ref{app:zeta}.
This is because in the absence of dust $\kappa$ is 
a function not only of gas temperature but also of gas 
density. As demonstrated in R06 in this case $M_{crit}$ is no 
longer independent of the ambient temperature $T_0$ and density 
$\rho_0$ of the nebula as the original analyses of Mizuno (1980)
and Stevenson (1982) suggest. Instead one finds that 
$M_{atm}\propto (\rho_0/T_0^3)^{q/(1+q)}$, where 
$q\equiv \partial\ln\kappa/\partial\ln\rho$ (R06). Obviously, $M_{crit}$
then also depend on $\rho_0$ and $T_0$, and the knowledge of 
this dependence is very important for obtaining $\Sigma_{lim}$ 
and $a_{lim}$ in the dust-free case. Unfortunately, we do not 
possess this knowledge from first principles as opacity 
calculations are rather complicated, and in any case 
we cannot currently calibrate $M_{crit}$ as functions of $M_c$, 
$T_0$ and $\rho_0$ against numerical results because calculations of 
Hori \& Ikoma (2010) were done for a single value of the planetary 
semi-major axis (meaning fixed values of $T_0$ and 
$\rho_0$), while $\dot M$ was varied. 
The scaling of $M_{atm}$ and $M_{crit}$ with $T_0$ and $\rho_0$ 
and its implication for the possibility of the CI thus remain worthwhile 
issues for future investigation.

We can still get a qualitative idea of how $a_{lim}$ changes in 
the dust-free case by setting opacity in equation (\ref{eq:a_lim}) 
at the very low level consistent with pure gas opacity, e.g.
$\kappa=10^{-4}$ cm$^2$ g$^{-1}$. We then find 
$a_{lim}^{MMSN}\approx 80$ AU compared to $44$ AU that equation 
(\ref{eq:a_lim}) predicts for $\kappa=0.1$ cm$^2$ g$^{-1}$.
Thus, opacity reduction due to sedimentation and coagulation 
of dust grains in the protoplanetary atmosphere may
help in extending the range of distances in which the CI is possible.

\section{Discussion.}
\label{sect:disc}

Despite the robustness of our arguments
it is not inconceivable that some additional factors can weaken
them and make giant planet formation by the CI possible
even beyond the limits represented by equations (\ref{eq:Sig_lim}) 
and (\ref{eq:a_lim}). Alternatively, it is quite possible that 
some of the assumptions used in deriving these results are too 
extreme and one can get even better constraints by focussing on 
less dramatic assumptions. Below we review factors that 
can work one way or another.

\subsection{Extending CI to larger radii.}
\label{subsect:weaken}

One possible way to  facilitate CI and increase $a_{lim}$ is to
consider possibility of planetesimal accretion at rates {\it exceeding}
$\dot M_{max}$. This is very difficult (since there are many factors 
that tend to reduce $\dot M$ compared to $\dot M_{max}$, see 
\S\ref{subsect:strengthen}) but may be possible if e.g. one takes into 
account the increase of planetesimal capture cross-section by 
the core caused by its extended, dense atmosphere. This effect has
been previously investigated by Inaba \& Ikoma (2003) who 
demonstrated that an increase of $\dot M$ by a factor 
of $\sim 10$ compared to the value computed without atmosphere 
is possible. According to equation (\ref{eq:a_lim}) such an 
enhancement of $\dot M$ (incorporated by increasing $\chi$) would boost 
$a_{lim}^{MMSN}$ by a factor of $\sim 4$.

Our present calculations assume that the core is accreting planetesimals 
continuously until the protoplanetary nebula dissipates --- this is 
important at large $a$ since massive core requires long time to be built.
But one may wonder if building smaller core in shorter time and then 
cutting off subsequent planetesimal accretion (and energy release 
at the core surface, which supports atmosphere against going unstable) 
completely may still lead to the CI and potentially extend it to 
larger semi-major axes. Such accretion scenario has been adopted 
by e.g. Pollack \etal (1996). The problem in this case is that 
even if $\dot M=0$ it still takes long time for the atmosphere around
the core to grow to the mass comparable to $M_c$. INE0 show that this
process occurs on thermal timescale of the atmosphere and typically 
takes millions of years. 

Similar problem is also encountered in a scenario where the core
grows rapidly by planetesimal accretion in the inner regions of 
protoplanetary disk and then gets scattered out to large radii by 
some massive perturber. One might expect that after the orbit of the scattered 
core circularizes by dynamical friction the core would gradually accrete 
massive atmosphere and undergo CI at some point. Given that both the 
orbit circularization and envelope accretion are likely to take long 
time it is not at all obvious whether the CI could be achieved in 
this scenario within several Myrs.

\subsection{Limiting CI to smaller radii.}
\label{subsect:strengthen}

There are many factors that can potentially reduce $a_{lim}$
compared to 44 AU estimated in equation (\ref{eq:a_lim}). For example, 
there are several reasons why it may not be possible for 
$\dot M$ to reach the maximum rate $\dot M_{max}$. 

First, the growing core 
may clear out a gap in planetesimal disk around its orbit, thus
significantly 
reducing $\dot M$ (Tanaka \& Ida 1997; Rafikov 2001; 2003a). 
In our previous calculations we implicitly assumed 
this not to happen e.g. because of the core migration through
the disk, which allows fresh planetesimal material to be constantly
supplied for core accretion (Alibert \etal 2005). 

Second, as we mentioned in
\S\ref{sect:accr}, a known pathway to $\dot M\approx\dot M_{max}$ is
via the growth of the core to the size at which it starts dominating 
dynamical evolution of nearby planetesimals and triggers 
their efficient collisional fragmentation (Rafikov 2004). 
However, there is a strong implicit assumption in this scenario ---
that the core can reach this critical size within the nebula lifetime.
Rafikov (2003b) has shown that at $a\sim 30-40$ AU a dynamically dominant 
core would need to have mass of order $10^{24}$ g and would require on
the order of $10-100$ Myr to grow in the MMSN. This time scale is 
apparently in conflict with the typical dissipation times of 
protoplanetary disks. Thus, one may need to either require a more  
massive disk at these radii or to find other pathways for accretion at 
$\dot M_{max}$. Formation of massive solid bodies by direct gravitational
instability facilitated by various streaming instabilities 
(Johansen \etal 2009) may be quite relevant for the latter option. 
Regardless of this (arguably rather serious) issue our estimate 
(\ref{eq:a_lim}) still remains a useful upper limit on $a_{lim}$. 

Among other factors hindering the onset of CI and reducing 
$a_{lim}$ we should mention the possibility of high opacity in the 
protoplanetary atmosphere. It was suggested in \S\ref{subsect:opacity} 
that $\kappa$ may be very low because of the dust sedimentation
and growth. However, infalling planetesimals which feed core accretion 
likely get partly disrupted in the atmosphere leaving behind large amount
of refractory material. This may actually {\it increase} $\kappa$ compared
to the value of $0.1$ cm$^2$ g$^{-1}$ assumed in 
equation (\ref{eq:a_lim}). Nevertheless, given the weak sensitivity of 
$a_{lim}$ to $\kappa$ the potential increase of atmospheric opacity 
is unlikely to have huge effect on $a_{lim}$.

\section{Summary.}
\label{sect:sum}

We studied the formation of giant planets by core instability at 
different locations in the protoplanetary disk with the goal of 
determining the range of radii where the CI is feasible  
within the several Myr lifetime of the protoplanetary 
disk. We demonstrate that this range is determined by two key 
factors: 
\begin{itemize}

\item The high planetesimal accretion rate is necessary to build 
the solid core as rapidly as possible at large separations 
from the star.

\item Intense energy release at the core surface caused by 
planetesimal accretion increases the critical core mass and
delays the CI.

\end{itemize}
The first factor turns out to be more important and the largest
distance at which CI can happen, around 40-50 AU, is obtained when 
$\dot M$ is at its highest possible value corresponding to 
accretion of dynamically cold planetesimals. The core mass 
corresponding to this case is around $15$ M$_\oplus$, likely 
compatible with the isolation mass at this distance.  

Our approach is quite different from other similar studies which 
often assume that (1) accretion proceeds at much slower rate $\dot M_{tr}$ 
defined by equation (\ref{eq:M_tr}) corresponding to accretion of
planetesimals moving with random velocities at the level of $\Omega R_H$ 
and/or (2) the CI commences after the core has reached a fixed mass of
around 10 M$_\oplus$ irrespective of the planetesimal accretion rate
or the location in the disk. Relaxing these two arbitrary assumptions we 
are able to obtain a significantly more robust and self-consistent 
limit on the CI operation which can be represented as a lower 
bound on the solid surface density ($\gtrsim 0.1$ g cm$^{-2}$)
or an upper bound on the size of the region where 
the CI can get going within several Myr timescale. These limits
are insensitive to the mass of the central star and depend only
weakly on the opacity in the core atmosphere. Our predictions are
relevant for interpreting the results of current and future direct 
imaging surveys (Marois \etal 2008) designed to uncover and 
characterize the population of giant planets at large separations 
from their parent stars.

\acknowledgements

Author is grateful to Ruth Murray-Clay for useful discussion and 
the suggestion of the limiting radius calculation for the marginally 
gravitationally unstable disk. The financial support for this work 
is provided by the Sloan Foundation and NASA via grant NNX08AH87G.


\appendix

\section{Calibration of $\zeta$.}
\label{app:zeta}

INE0 investigated the dependence of $M_{crit}$
on $\kappa_0$ and $\dot M$ assuming the latter to be constant. 
Their numerical calculations of the CI include 
processes such as the dissociation 
and ionization of hydrogen, realistic gas opacities with 
the inclusion of the dust grain contribution, and self-gravity 
of the atmosphere. These are the crucial ingredients missing
in the analytical models of Stevenson (1982) and R06. Thus,
it is expected that calculations of INE0 should be more realistic 
than the aforementioned analytical studies. 

We calibrate the coefficient $\zeta$ entering 
the expression for $M_{atm}$ against the results of INE0 who found
that for $\kappa_0\gtrsim 10^{-2}$ cm$^2$ 
g$^{-1}$ (see equation A2 in INE0)
\ba
M_{crit}\approx 7\, M_\oplus\left(\frac{\dot M}{10^{-7}
\,M_\oplus\,\mbox{yr}^{-1}}\right)^{q^\prime}
\left(\frac{\kappa_0}{1
\,\mbox{cm}^2\,\mbox{g}^{-1}}\right)^{s},
\label{eq:fit}
\ea
and it follows from their Table 6 that the best-fit power law 
exponents $q^\prime$ and $s$ satisfy $q^\prime = 0.27-0.29$
and $s=0.24-0.29$. Motivated by these scalings we first assume 
power law dependence in the form
\ba
\zeta=\zeta_0\eta M_c^{\delta} L^{-\omega}. 
\ea
The explicit dependence of $\zeta$ on parameter $\eta$ is motivated by the
core instability condition (\ref{eq:instab_cond}) and the assumed 
independence of $\eta$ on the planetesimal accretion history 
(i.e. the same value of $\eta$ applies in our case of $\dot M\propto 
M_c^{2/3}$, see eq. [\ref{eq:mdot}],  as in the 
$\dot M=$const case studied by INE0). Plugging this expression 
into (\ref{eq:Matm}) and using $L=GM_c \dot M/R_c$ and equation 
(\ref{eq:instab_cond}) we find
\ba
M_{crit}=\left[\frac{\left(k_B/\mu\right)^4}{\zeta_0\sigma G^{3-\omega}}
\left(\frac{4\pi\rho_c}{3}\right)^{(1+\omega)/3}\right]^{\tilde s}
\kappa_0^{\tilde s}\dot M^{\tilde q},
\label{eq:M1}
\ea
where 
\ba
\tilde q=\frac{3(1+\omega)}{7+3\delta-2\omega},~~~~~
\tilde s=\frac{3}{7+3\delta-2\omega},
\label{eq:exps1}
\ea
play the role of $q^\prime$ and $s$ in INE0 case. It follows 
from (\ref{eq:exps1}) that $\tilde q/\tilde s=1+\omega$
while the INE0's results for these indices give $1\lesssim 
q^\prime/s\lesssim 1.2$.
This constrains $\omega \lesssim 0.2$ resulting in a very weak dependence 
of $\zeta$ on $L$. Based on this observation and given the 
approximate nature of our calibration 
procedure we decided to neglect the dependence of $\zeta$ on $L$ altogether
and to consider $\zeta$ in the form (\ref{eq:zeta}), i.e. scaling with 
$M_c$ only. Repeating our calculation we find instead of (\ref{eq:M1})
the following expression for $M_{crit}$:
\ba
M_{crit}=\left[\frac{\left(k_B/\mu\right)^4}{\zeta_0\sigma G^3}
\left(\frac{4\pi\rho_c}{3}\right)^{1/3}\right]^{\tilde s}
\kappa_0^{\tilde s}\dot M^{\tilde q},~~~~~~\tilde s=\tilde q=
\frac{3}{7+3\delta}.
\label{eq:M2}
\ea
Choosing $\delta=1.2$ as stated in equation (\ref{eq:zeta}) we find 
$\tilde s=\tilde q\approx 0.28$ in good agreement with the $q^\prime$ 
and $s$ values found in INE0. 

To determine $\zeta_0$ we use the fact that according to equation 
(\ref{eq:fit}) $M_{crit}\approx 7$ M$_\oplus$ for $\kappa_0=1$ cm$^2$ 
g$^{-1}$ and $\dot M=10^{-7}$ M$_\oplus$ yr$^{-1}$. This uniquely 
determines the value of $\zeta_0$ in equation (\ref{eq:zeta}).

Note that in carrying out this calibration we implicitly used the
fact that $M_{crit}$ is independent of the ambient conditions in 
the nebula if $\kappa$ is independent of gas density (R06), which is true 
if opacity is independent of gas density. This is an
important point since the INE0 calculations have been done 
without varying the external conditions. More on this issue can
be found in \S\ref{subsect:opacity}.

\end{document}